\documentclass[AMA,STIX1COL]{WileyNJD-v2}

\usepackage{hyperref}
\hypersetup{
  colorlinks = true, 
  citecolor = blue, 
  linkcolor = blue, 
  urlcolor = black
}

\usepackage{graphicx}
\usepackage{amsmath}
\usepackage{mathtools}
\usepackage{pifont}
\usepackage{xcolor}
\usepackage{float}
\usepackage{url}
\usepackage{tabularx}
\usepackage{slashbox}
\usepackage{algorithm}
\usepackage{algorithmicx}
\usepackage{algpseudocode}
\usepackage{setspace}
\usepackage[compact]{titlesec}

\definecolor{grey}{HTML}{757575}

\newcommand\Tstrut{\rule{0pt}{2.6ex}}
\newcommand\Bstrut{\rule[-0.9ex]{0pt}{0pt}}

\newcommand{\rw}[1]{\textcolor{black}{#1}}
\newcommand{\wy}[1]{\textcolor{black}{#1}}
\newcommand{\ct}[1]{\textcolor{black}{#1}}
\newcommand{\rev}[1]{\textcolor{black}{#1}}
\newcommand{\revision}[1]{\textcolor{black}{#1}}

\articletype{RESEARCH ARTICLE}

\received{}
\revised{}
\accepted{}
        
\begin{document}


\title{How does SSD Cluster Perform for Distributed File Systems: An Empirical Study}


\author[1,2]{Jiashu Wu}
\author[1]{Yang Wang*}
\author[1,3]{Jinpeng Wang}
\author[1,3]{Hekang Wang}
\author[4]{Taorui Lin}

\authormark{Jiashu Wu \textsc{et al}}

\address[1]{\orgname{Shenzhen Institute of Advanced Technology, Chinese Academy of Sciences}, \orgaddress{\state{Shenzhen 518055}, \country{China}}}
\address[2]{\orgname{University of Chinese Academy of Sciences}, \orgaddress{\state{Beijing 100049}, \country{China}}}
\address[3]{\orgname{University of Science and Technology of China}, \orgaddress{\state{Hefei 230026}, \country{China}}}
\address[4]{\orgname{Shenzhen Virtual Clusters Information Technology Co., Ltd}, \orgaddress{\state{Shenzhen 518057}, \country{China}}}

\corres{*Corresponding Author: Yang Wang, \email{yang.wang1@siat.ac.cn}}

\presentaddress{1068 Xueyuan Avenue, Shenzhen University Town, Shenzhen 518055, Guangdong, P.R.China}


\abstract[Summary]{As the capacity of Solid-State Drives (SSDs) is constantly being optimised and boosted with gradually reduced cost, the SSD cluster is now widely deployed as part of the hybrid storage system in various scenarios such as cloud computing and big data processing. However, despite its rapid developments, the performance of the SSD cluster remains largely under-investigated, leaving its sub-optimal applications in reality. To address this issue, in this paper we conduct extensive empirical studies for a comprehensive understanding of the SSD cluster in diverse settings. To this end, we configure a real SSD cluster \revision{and gather} the generated trace data based on some often-used benchmarks, then adopt analytical methods to analyse the performance of the SSD cluster with different configurations. \revision{In particular, regression models are built to provide better performance predictability under broader configurations}, and the correlations between influential factors and performance metrics with respect to different numbers of nodes are investigated, which reveal the high scalability of the SSD cluster. Additionally, the cluster's network bandwidth is inspected to explain the performance bottleneck. Finally, the knowledge gained is summarised to benefit the SSD cluster deployment in practice. }


\maketitle


\section{Introduction}\label{sec:section1}

With rapid developments of technologies such as cloud computing\textsuperscript{\cite{pan2017future}} and big data analytics\textsuperscript{\cite{wu2022toward,li2020simultaneous,10026337}}, efficient and reliable storage technologies with high throughput and scalability become increasingly indispensable and thus are drawing great attention from both industry and academic communities. Among various storage mediums, the Solid-State Drive (SSD), by virtue of its high performance, has been identified as a mainstream substrate for storage systems. 

\ct{Unlike Hard Disk Drives (HDDs), SSDs are composed of semiconductor chips and thereby providing exceptional performance compared to HDDs\textsuperscript{\cite{chakraborttii2020improving}}. On the other hand, as the cost of the SSD keeps dropping\textsuperscript{\cite{wei2015z_ssd_price_down}}, it is becoming more affordable to deploy SSDs in a cluster form in the storage systems. While the SSD is ideal for performance optimisation, completely replacing HDDs with SSDs in large-scale storage system is not a widely adopted solution due to the concern of the monetary cost (compared to HDDs), the features of SSD, and other factors. \revision{Thus, a hybrid DFS, which exploits a bunch of SSDs to work as a \emph{small} cluster (i.e., the SSD cluster in our particular sense) to facilitate the storage system as a whole is more practical in reality\textsuperscript{\cite{zhou2020efficient,wadhwa2019iez}}. For example, the SSD cluster is in practice often used to optimise the storage of small data\textsuperscript{\cite{Marc2020GekkoFS_small_metadata,10.1145/3490237}}, metadata\textsuperscript{\cite{7600359,8489897}} or functioning as caches for hot data\textsuperscript{\cite{9802898}} under common scenarios such as Internet of Things (IoT)\textsuperscript{\cite{9833301,9933783}}. }}

Undoubtedly, as technologies such as big data analytics and cloud computing keep evolving and the amount of data grows up exponentially\textsuperscript{\cite{li2021self_big_data,DAI2022108}}, the SSD cluster would become more crucially important and broadly used. In recent years, the SSD's capacity has been as expected to substantially boost\textsuperscript{\cite{kang2018subpage_ssd_capacity_increase,wang2019project_ssd_capacity_increase_high_performance}}, which makes \ct{the SSD cluster, originally designed for some special purposes,} an \ct{attractive choice for general uses in} \wy{those} applications that require high throughput, reliability and scalability \ct{as by its way, a type of storage cluster can be constructed effectively with multi-level configuration for SSDs to serve user's diverse needs. }

\rev{Typically, the advocated SSD cluster consists of \wy{a \emph{small} group of} storage nodes, each being equipped with \wy{a bunch of} SSDs. }As a result, this hierarchical structure not only allows the storage cluster to inherit the superiority of the SSD, such as wonderful throughput and low latency, but also enables it to provide reliability and scalability warranties, benefitted from the mechanisms of storage clusters such as resource duplication. Currently, numerous applications have testified to the promising performance of the SSD cluster. For example, Google Clouds\textsuperscript{\cite{google_clouds1,google_clouds2}} enables the SSD cluster to be used as part of the storage system of its cloud services, demonstrating SSD cluster's excellent throughput performance and scalability\textsuperscript{\cite{koo2015dual_SSD_cluster_advantage}}.

Despite rapid developments and prevalent applications of the SSD cluster, compared to a single SSD \ct{whose performance has been studied intensively in recent years \textsuperscript{\cite{cheong2012ddr_cheong_related_work,son2015empirical_son_related_work}},} \rev{the study on the performance of the SSD cluster is still quite few and far between,} making its performance remain under-investigated. \rw{However, for the successful deployments of the SSD cluster with respect to the distributed file system (DFS), it is essential to understand its performance-critical factors in various settings. Thus, the answers to the following questions are valuable to gain better understanding of the characteristics of the SSD cluster, and therefore provide insights and guidances to the deployment of SSD clusters: }
\rw{\begin{itemize}
    \item How does the SSD cluster perform \ct{with respect to} different interface types, different I/O access patterns, different number of client processes and different data chunk sizes?
    \item \revision{Whether or not the performance behaviours of the SSD clusters are predictable across different settings? }
    \item Whether or not the performance influential factors present stable influence on performances of the SSD cluster when the number of nodes in the cluster varies?
\end{itemize}}
\rw{Driven by the SSD cluster's increasingly wide deployment and its under-examined performance, it naturally leads to this research, in which the performance of the SSD cluster \revision{in a relatively small size} is comprehensively evaluated. }
\wy{To this end, a real SSD cluster is configured and a Ceph-based DFS is set up to collect the trace data based on some often-used benchmarks \ct{with an attempt to make answers to above questions} persuasive and practical. Then, some analytical methods are leveraged to comprehensively analyse the performance of the SSD cluster in various settings.}

\rw{Briefly, the contributions of this paper are summarised as follows: 
  \begin{itemize}
    \item \revision{We developed a small-scale prototype cluster where a Ceph-based DFS is deployed to} evaluate the performance of the SSD cluster \ct{in} various settings (such as number of nodes, SSD interface types, I/O access patterns, number of client processes and data chunk sizes) and reported the experimental results with respect to several often-used evaluation metrics. 
    \item \ct{We comprehensively analysed the evaluation results using some statistical tools, \revision{which can make the results not only more concrete, but also more generalisable and practical. }}
    \item \revision{We answered the above research questions regarding the SSD cluster from several important aspects --- the performance, predictability, stability and scalability in different settings.} We provided several lessons and knowledge gained from the performance evaluation which are useful to guide deployments of SSD clusters in future. 
\end{itemize}}

The rest of this paper is organised as follows: Section \ref{sec:section2} introduces technical backgrounds of the SSD cluster in a nutshell, which includes SSD's mechanism and the structure of the SSD cluster that forms part of the file system. Related works are discussed in Section \ref{sec:section3}. Section \ref{sec:section4} presents the experimental setup. \rev{Results are then comprehensively analysed and discussed in Section \ref{sec:section5} with lessons and knowledge being provided. }Section \ref{sec:section6} concludes the paper.


\section{SSD Cluster Technical Background}\label{sec:section2}

As the amount of data being generated grows rapidly\textsuperscript{\cite{waller2013data_big_data_grow_exponentially,li2021self_big_data}} and key applications become more data-intensive\textsuperscript{\cite{8616889}}, efficient storage technologies such as the Solid-State Drive (SSD) and the storage cluster architecture gradually gain prominence. This section provides some technical details of the SSD. 

\subsection{Solid-State Drive}\label{sec:section2.1}

The SSD is now becoming a promising storage technology. We now introduce some internal mechanisms\textsuperscript{\cite{ssd_technical_background}} of the SSD relevant to the experiments in detail. 

A Solid-State Drive is a flash-memory based data storage device. Voltages are applied to store bits into cells made by floating-gate transistors, and NAND-flash memory is a popular solution of how bits are being read, written, and erased. An important property of NAND-flash memory is that it has a limited lifespan. Its cells will be worn off when a certain limit of Program/Erase (P/E) cycle is reached. 
Due to the organisation of NAND-flash cells, read, write, and erase operations cannot be performed on a single cell individually. Cells are grouped into pages, and read/write operations can only be done at the page level. Reads that are not aligned on the multiply of page size will end up reading more than necessary, same for writes as well. Even though a write operation affects only a single bit, a whole page will be written anyway, which is known as write amplification. And that is the reason why some data chunk size settings are better than others. Moreover, pages cannot be overwritten in the NAND-flash memory, old pages will be flagged as being staled and wait to be garbage collected, causing the read-modify-write operation. 
That explains why in SSDs, writes are more inefficient than read operations. Finally, erase operations are performed at the block level, where a block is formed by multiple pages, and it is not possible to erase an individual page. When write and erase operations are not distributed evenly among all cells, some cells will be worn off earlier than others, causing the capacity of the SSD to drop gradually. Hence, SSD manufacturers implement the wear-leveling mechanism in the SSD's controller 
to balance workloads between cells so that cells will reach their P/E cycle limit and wear off roughly at the same time to maintain the capacity and lifetime of the SSD as much as possible.

Another mechanism worth mentioning is the garbage collection implemented in the SSD's controller. It is in charge of erasing staled pages and restoring its state into ``free''. 
Furthermore, over-provisioning is proposed to use the reserved blocks to act as a buffer under heavy workloads which leaves the garbage collector more time to perform the time-consuming erase operations. Hence, the overall performance of the SSD can be optimised. 

The hierarchical structure of the NAND-flash memory 
makes parallelism a perfect choice to boost SSD's performance. Hence, SSD provides channel-level, package-level, chip-level and plane-level parallelism, aiming to optimise the performance as much as possible. And that explains why using a reasonable number of parallel threads when interacting with the SSD is a good choice. Also, thanks to the internal parallelism of SSDs, when fully exploited, random and sequential operations show similar performance, which boosts the SSD's efficiency.


\section{Related Work}\label{sec:section3}

As the SSD keeps evolving and its capacity becomes increasingly promising\textsuperscript{\cite{kang2018subpage_ssd_capacity_increase,kang2014multi_ssd_capacity_increase_high_performance,wang2019project_ssd_capacity_increase_high_performance}}, its applications become more prevalent and viable in storage systems\textsuperscript{\cite{micheloni2017solid_ssd_broad_applications}}. With the radical changes it brings to the storage systems, the SSD has attracted more attention by both industry and academic communities\textsubscript{\cite{7805265}}. Several past research efforts have been focusing on the performance evaluation of SSDs. 

Cheong et al.,\textsuperscript{\cite{cheong2012ddr_cheong_related_work}} conducted the SSD performance evaluation with varied number of clients connected with a single SSD storage node in a unified storage system for cloud computing. Two performance metrics including the total execution time and transactions per second had been evaluated. However, this work was relatively limited as only two performance metrics were tested and evaluated, lacking several important performance metrics. Moreover, the work only considered one variation factor, i.e., the client-storage ratio, which impaired the comprehensiveness of this work. Son et al.,\textsuperscript{\cite{son2015empirical_son_related_work}} considered NVM SSD's high throughput and scalability attractive for storage systems and hence evaluated the performance of the NVM SSD. Performance metrics such as throughput had been reported and analysed under different I/O access patterns and different number of threads. Dell's lab\textsuperscript{\cite{kasavajhala2011solid_dell_related_work}} also put effort on the evaluation of storage device's performance. Their work compared the bandwidth and IOPS between the Hard Disk Drive (HDD) and SSD. However, these two works only focused on limited performance metrics, and key performance metric including latency was absented. Statistical tools are beneficial on the evaluation of performance data, however, they all failed to utilise them when performing the analyses. What's more, only one or two performance influential factors were varied, which hurt the completeness of these works. 

Ahn et al.,\textsuperscript{\cite{ahn2015analytical_ahn_related_work}} evaluated the SSD under MapReduce workloads, with the discussion of the queuing network which can be leveraged to predict and simulate the execution time of MapReduce jobs. Kim et al.,\textsuperscript{\cite{kim2009flashsim_kim_related_work}} proposed a simulator called FlashSim, which can provide a simulation environment for NAND flash-based SSDs. They then testified to the effectiveness of the proposed simulator by using it to evaluate the energy consumption and average system response time of the SSD. 

Unfortunately, all of the aforementioned works attempted to conduct the performance evaluation of a single SSD storage device or node, and none of them put their effort to the evaluation of SSD cluster's performance. The performance evaluation of a single SSD is not sufficient to reveal the performance characteristics of the SSD cluster, and hence making the performance evaluation of SSD clusters a void that needs to be filled. 

Recently, Jeremic et al.,\textsuperscript{\cite{jeremic2011pitfalls_jeremic_related_work}} evaluated the performance of multiple SSD nodes using IOPS and normalised random write performance as their performance evaluation metrics. However, their evaluation still suffered from the lack of comprehensiveness, as important performance influential factors such as different number of client processes were not evaluated at all and the characteristics of the SSD cluster such as scalability were not fully explored. 

The SSD has been in the storage market for a relatively short time\textsuperscript{\cite{kim2009flashsim_kim_related_work}}, despite there were some previous efforts in the evaluation of SSD's performance, to our best knowledge, there still lacks a comprehensive performance evaluation of the SSD cluster. Hence, we aim to fill the void by evaluating three important performance metrics of the SSD cluster under several different performance influential factor settings, which include different SSD interface types, different number of storage nodes in the SSD cluster, different I/O access patterns, different data chunk sizes and different number of processes per client. We believe our evaluation will be beneficial and will contribute towards better understanding and applications of SSD clusters.


\section{Experimental Setup}\label{sec:section4}

In this section, we will introduce the experimental setup \revision{from the following perspectives}: SSD cluster architecture, hardware and file system configurations, benchmarking tools, evaluation metrics and experimental settings. 

\begin{figure*}[!ht]
  \begin{center}
    \includegraphics[width=0.7\textwidth,keepaspectratio]{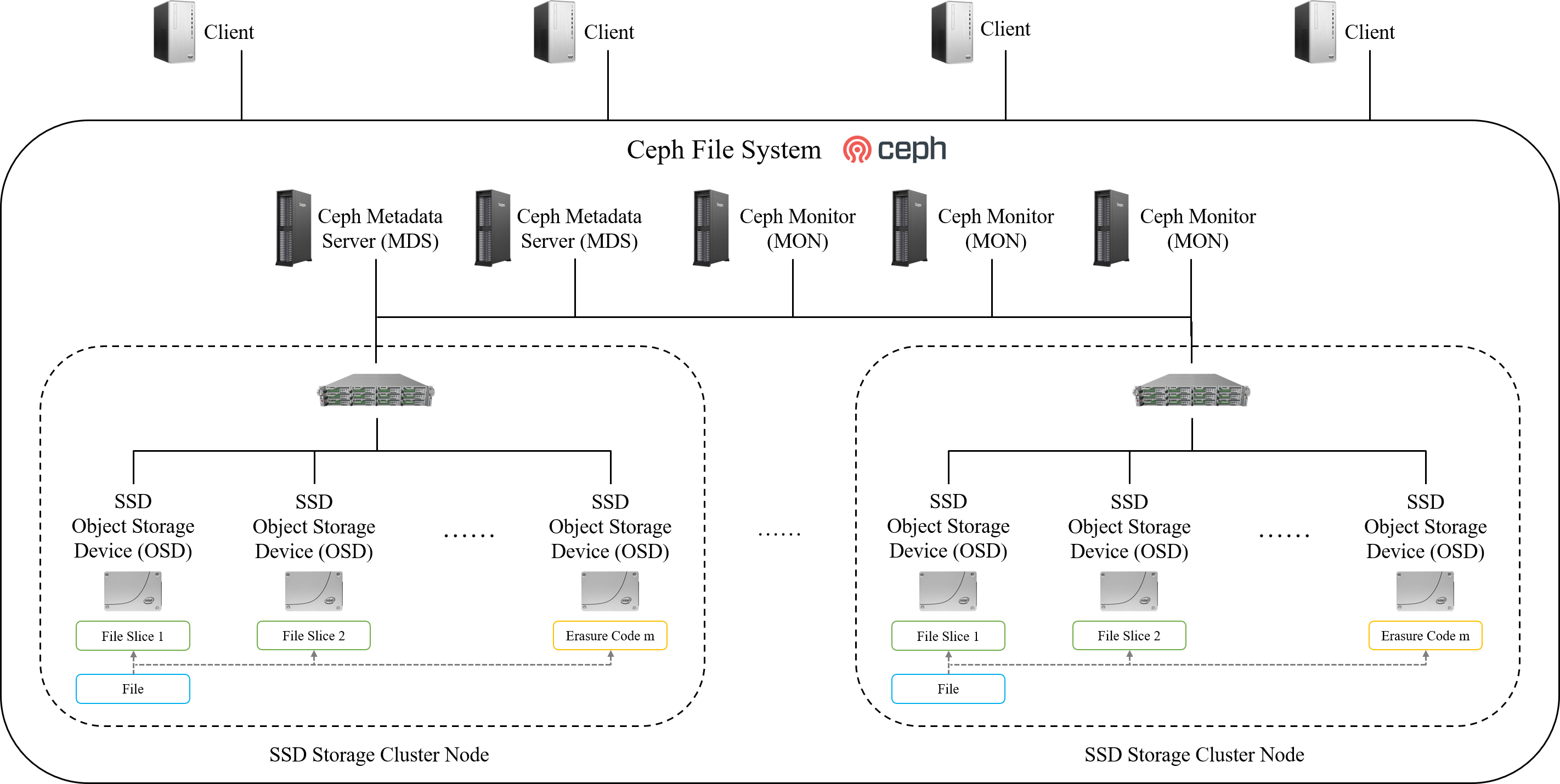}\\
    \caption{The structure of the SSD cluster used during experiments}
    \label{fig:figure2}
  \end{center}
\end{figure*}

\subsection{SSD Cluster Architecture}\label{sec:section2.2}

The SSD cluster in this paper forms part of the Ceph file system\textsuperscript{\cite{weil2006ceph_ceph_file_systems}}, a file system that's commonly used in distributed systems and cloud computing systems. The architecture is illustrated in Figure \ref{fig:figure2}. \revision{Again, the current SSD cluster typically acts as part of the hybrid DFS to store frequently modified data due to its excellent performance. Hence, the SSD cluster is usually deployed in small-scale to facilitate the storage system as a whole, and is especially useful to store small data, metadata or hot data for scenarios such as the IoT scenario. However, the performance evaluation specifically targeting SSD cluster with relatively small-scale remains under-investigated. Considering this, a SSD cluster with relatively small scale is leveraged during our experiment to guide future deployments of SSD clusters in general usage. }

The Ceph file system consists of Ceph Metadata Servers (MDS), Ceph Monitors (MON), and Object Storage Devices (OSD). The MDS manages file metadata of the storage system. The MON maintains a master copy of the cluster map to ensure high availability of the SSD cluster and hence benefit its reliability and scalability. The OSD in our case is a server equipped with multiple SSDs as shown in dotted boxes in Figure \ref{fig:figure2}, which stores the data for the file system. Our SSD cluster is highly scalable, the number of SSDs in the cluster can be flexibly adjusted. Furthermore, we utilise several OSDs, hence forming a "two-level" SSD cluster, i.e., several storage nodes, with multiple SSDs in each node. The SSD cluster we utilised distributes data as evenly as possible among nodes to avoid read/write hotspots, so that the throughput performance can be highly efficient. To make the SSD cluster more reliable, we use the erasure code mechanism, which slices the file into n partitions with m erasure codes. They are then stored into different SSDs in the cluster so that the SSD cluster is operable so long as there are less than m failed nodes. 

As drawn on top of the Figure \ref{fig:figure2}, the client devices interact with the storage system via network connections, sending various requests under different settings including different I/O access patterns, different data chunk sizes, etc. 

\begin{table}
  \centering
  \caption{Device configuration of server nodes and clients}
  \vspace{3mm}
  \setlength{\tabcolsep}{2.0mm}{
    \begin{tabularx}{0.8\textwidth}{c|llll}
      \toprule
      Device Type & Brand & Processor & Memory & OS \\
      \hline
      Server node & Sugon & Intel Xeon CPU E5-2630 V3 @ 2.40 GHz & 64 GB & CentOS 7.2.1511 \\[0.3em]
      Client & HP & Intel Xeon L5520 @ 2.26 GHz & 8GB & CentOS 7.2.1511 \\
      \bottomrule
    \end{tabularx}
  }
  \label{tab:table1}
\end{table}

\begin{table}
  \centering
  \caption{SSD model and specification}
  \vspace{3mm}
  \setlength{\tabcolsep}{2.0mm}{
    \begin{tabularx}{0.65\textwidth}{c|lll}
      \toprule
      SSD Type & Model & Capacity & Interface Type\Bstrut \\
      \hline
      SATA SSD & Intel 5400S & 240 GB & SATA 3.0\Tstrut \\[0.3em]
      PCIe SSD & Shannon Direct-IO PCIe Flash-G3i & 1.6 TB & PCIe 2.0\Bstrut \\
      \bottomrule
    \end{tabularx}
  }
  \label{tab:table2}
\end{table}

\subsection{Hardware and File System Configuration}\label{sec:section4.1}

We conduct our evaluations using the SSD cluster with architecture illustrated in Figure \ref{fig:figure2}. The file system we utilised is Ceph\textsuperscript{\cite{weil2006ceph_ceph_file_systems}}, in which we configure $2$ servers as the Ceph Metadata Server (MDS), as well as $3$ servers as the Ceph Monitor (MON). All these servers are Sugon servers\textsuperscript{\cite{sugon_server}} with their configuration details shown in Table \ref{tab:table1}. 

\rw{
Since the PCIe SSD cluster in general outperforms the SATA SSD cluster if they are equally sized \textsuperscript{\cite{krishnan2009integrating_pcie_outperform_sata}},
we deliberately resize the PCIe SSD cluster in such a way that its performance is comparable with the SATA SSD cluster. Thus, the Sugon server equipped with $6$ Intel SATA SSDs is leveraged, while for each PCIe SSD storage node, the Sugon server with $3$ Shannon PCIe SSDs is used to achieve a relatively fair play. The details of the SSD models are presented in Table \ref{tab:table2}. }

\rw{During experiments, multiple storage nodes, ranging from $2$ to $5$, are utilised as visualised in Figure \ref{fig:figure2}. The varied number of storage nodes can verify the scalability of the SSD cluster. Together, they form a SSD cluster and are managed by the Ceph file system, ready to process read and write requests sent by clients. The rationales of conducting performance evaluation on clusters with the aforementioned configuration are as follows: }
\rw{\begin{itemize}
    \item We follow previous works\textsuperscript{\cite{han2021empirical_small_cluster,cheong2012ddr_cheong_related_work}} to choose a SSD cluster with a relatively small size as the testbed, which consists of less than 8 nodes, each having less than 5 SSDs. Hence, our hardware configuration is comparable. 
    \item The SSD cluster we utilised during experiments is representative to process moderate I/O requests in a production environment\textsuperscript{\cite{vcluster_company}}\revision{, hence the results produced by the experiments are persuasive. The storage clusters we used are representatives of relatively small-scale storage clusters. }
    \item \rev{The scale of the cluster in our experiments is commmon to some typical experimental environments with 10 Gbps network since the SSD has a high throughput, leveraging a large SSD cluster with lots of storage nodes would quickly overwhelm the network. }
    \item \rev{Since the storage nodes in our SSD cluster are read/written in parallel, utilising more SSD storage nodes will yield a result that is relatively proportionally scaled up, assuming the network capacity is scaled accordingly. }
\end{itemize}}

As for clients, $8$ HP PCs with configuration shown in Table \ref{tab:table1} are used to send read and write requests to the SSD cluster via network connections. 

\rev{In terms of network environment, all servers are connected via a $10$ GBit/s network. }

\rev{Note that the hardwares we used during our experiments are usual commercial hardwares, and the file system that we used to organise the SSD storage cluster is Ceph, a file system that's commonly used in distributed systems and cloud computing systems. Hence, similar trends and lessons gained from the performance evaluation in this research should hold even when the evaluation is reproduced on other hardwares and system configurations. }

\begin{table}
  \centering
  \caption{Settings of performance influential factors}
  \vspace{3mm}
  \setlength{\tabcolsep}{2.0mm}{
    \begin{tabularx}{0.9\textwidth}{ll}
      \toprule
      Performance influential factors & Settings\Bstrut \\
      \hline
      Number of storage nodes in the SSD cluster & 2, 3, 4, 5\Tstrut \\[0.3em]
      SSD interface type & SATA, PCIe \\[0.3em]
      I/O access pattern & random read, sequential read, random write, sequential write \\[0.3em]
      Number of processes per client & 1, 8, 16 \\[0.3em]
      Data chunk size & 4 KB, 128 KB, 4 MB\Bstrut \\
      \bottomrule
    \end{tabularx}
  }
  \label{tab:table3}
\end{table}

\subsection{Benchmarking}\label{sec:section4.2}

\rw{In each client, the popular disk pressure testing tool Fio\textsuperscript{\cite{fio}} is used to generate workloads and send it from the client to the SSD cluster. The Fio \textsuperscript{\cite{fio,fio_overview,fio_linux}} is a popular open-sourced benchmarking tool that has been used to evaluate the performance of SSDs \textsuperscript{\cite{lee2019asynchronous,han2021empirical_small_cluster}}. It simulates a given I/O workload by spawning a number of threads or processes doing a particular type of I/O action as specified by the user. Hence it is a perfect tool to benchmark specific disk I/O workloads and is suitable in this case. It possesses various I/O APIs and allows users to define some threads or processes to submit works, in which the data chunk size, I/O type and data volume to be read/written can be specified to fit benchmarking needs. Hence, I/O workloads it generates are close to many server applications and can be used to simulate application requests sent from connecting clients in a generalised way. }

The size of the workload we used during our experiments is set to $32$ GB that simulates the workload of a normal production environment, rather than applying the unrealistic pre-conditioning workload that represents worst-case scenarios\textsuperscript{\cite{ssd_technical_background}}. 

\rw{The following settings are adjusted: 
  \begin{itemize}
    \item Data chunk size is varied among $4$ KB, $128$ KB and $4$ MB, to testify the SSD cluster's performance under small, medium and large data chunks, respectively. This can also verify the scalability of SSD clusters in terms of varied data chunk sizes. 
    \item The number of processes per client is adjusted within $1$, $8$, and $16$ to investigate the scalability of the SSD cluster when handling different number of request processes. 
    \item $4$ different I/O access patterns, including random read, sequential read, random write and sequential write, are used to test the SSD cluster's performance with respect to different I/O access patterns. 
    \item The comparisons between two interface types, i.e., SATA and PCIe, are performed. 
    \item The number of storage nodes in the SSD cluster is adjusted between $2$ and $5$ to inspect the SSD cluster's scalability. 
\end{itemize}}

Table \ref{tab:table3} summarises these different settings of performance influential factors. 

\begin{table}
  \centering
  \caption{Performance metrics used when evaluating SSD cluster's performance}
  \vspace{3mm}
  \setlength{\tabcolsep}{2.0mm}{
    \begin{tabularx}{\textwidth}{l|lp{7cm}l}
      \toprule
      Performance Metric & Abbreviation & Definition & Unit\Bstrut \\
      \hline
      Bandwidth & BW & Number of bits being transferred in one second & Mb/s\Tstrut \\[0.3em]
      Input/Output operations per second & IOPS & Number of input/output operations per second & Operation/s \\[0.3em]
      Complete latency & clat & Response time after the command is emitted till the completion & Milliseconds\Bstrut \\
      \bottomrule
    \end{tabularx}
  }
  \label{tab:table4}
\end{table}

\subsection{Evaluation Metrics}\label{sec:section4.3}

As for performance evaluation metrics, three important metrics are chosen to measure SSD cluster's performance, which include bandwidth, IOPS and complete latency. Their corresponding definitions have been presented in Table \ref{tab:table4}. Performance monitoring tool Nmon\textsuperscript{\cite{nmon}} is used to monitor the performance of the cluster and report the performance results. 

\rw{Nmon \textsuperscript{\cite{nmon,nmon_wiki,nmon_overview}} is a handy system monitoring tool that watches overall system performance and provides the user with a glance at system performance from many perspectives, such as CPU statistics, memory statistics, network statistics and errors, etc. The plentiful number of options it offers for system performance monitoring from different aspects make it a power solution for performance evaluation and data collection. Under online mode, it will display performance statistics on-screen in a easy-to-read format, while in offline mode, it will save these statistics to a CSV file for later plotting and processing to assist the understanding of system performances. }

\subsection{Experimental Settings}\label{sec:section4.4}

During the experiments, in order to make the results reproducible and plausible, experiments of each setting are repeated $30$ times. The parameter setting variations are presented in Table \ref{tab:table3}. 

Since each experiment is repeated 30 times, in all visualisations in this paper, the average experimental results are plotted with their corresponding error bars being presented.


\section{Result and Analysis}\label{sec:section5}

In this section, we present the evaluation results of the SSD cluster and the corresponding analyses. As the simplest SSD cluster, the performance of the 2-node SSD cluster will be presented first, followed by performance investigations of the SSD cluster that consists of multiple nodes. \revision{A regression model is built to enable better predictability on unseen configurations based on the rested results.} The correlation analysis is \revision{also} leveraged to analyse the scalability of the SSD cluster. \rw{Some lessons and knowledge learned from these analyses are given to guide further decision-making in cluster deployment and make our experiment concrete. }

\subsection{2-node SSD cluster performance evaluation and analysis}\label{sec:section5.1}

\begin{figure*}[!t]
  \begin{center}
    \includegraphics[width=\textwidth,keepaspectratio]{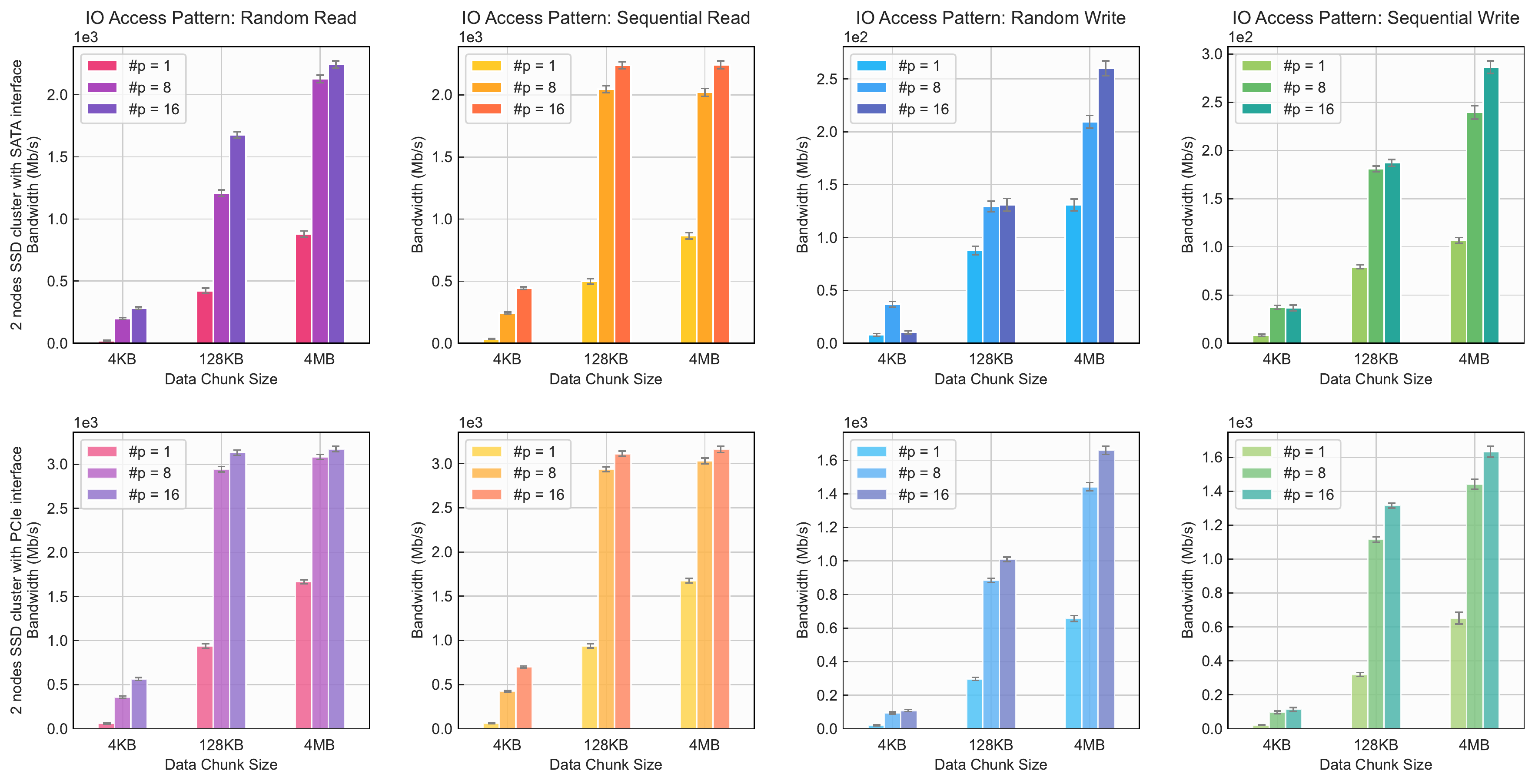}\\
    \caption{Bandwidth performance of the SSD cluster under different interface types (row in the figure), different I/O access patterns (column in the figure), different number of processes per client (represented in different colours), and different data chunk sizes (bar groups in the figure). $\#p$ represents number of client processes. \rw{The average results of 30 experiments are plotted, with corresponding error bars. }}
    \label{fig:figure3}
  \end{center}
\end{figure*}

\begin{figure*}[!ht]
  \begin{center}
    \includegraphics[width=\textwidth,keepaspectratio]{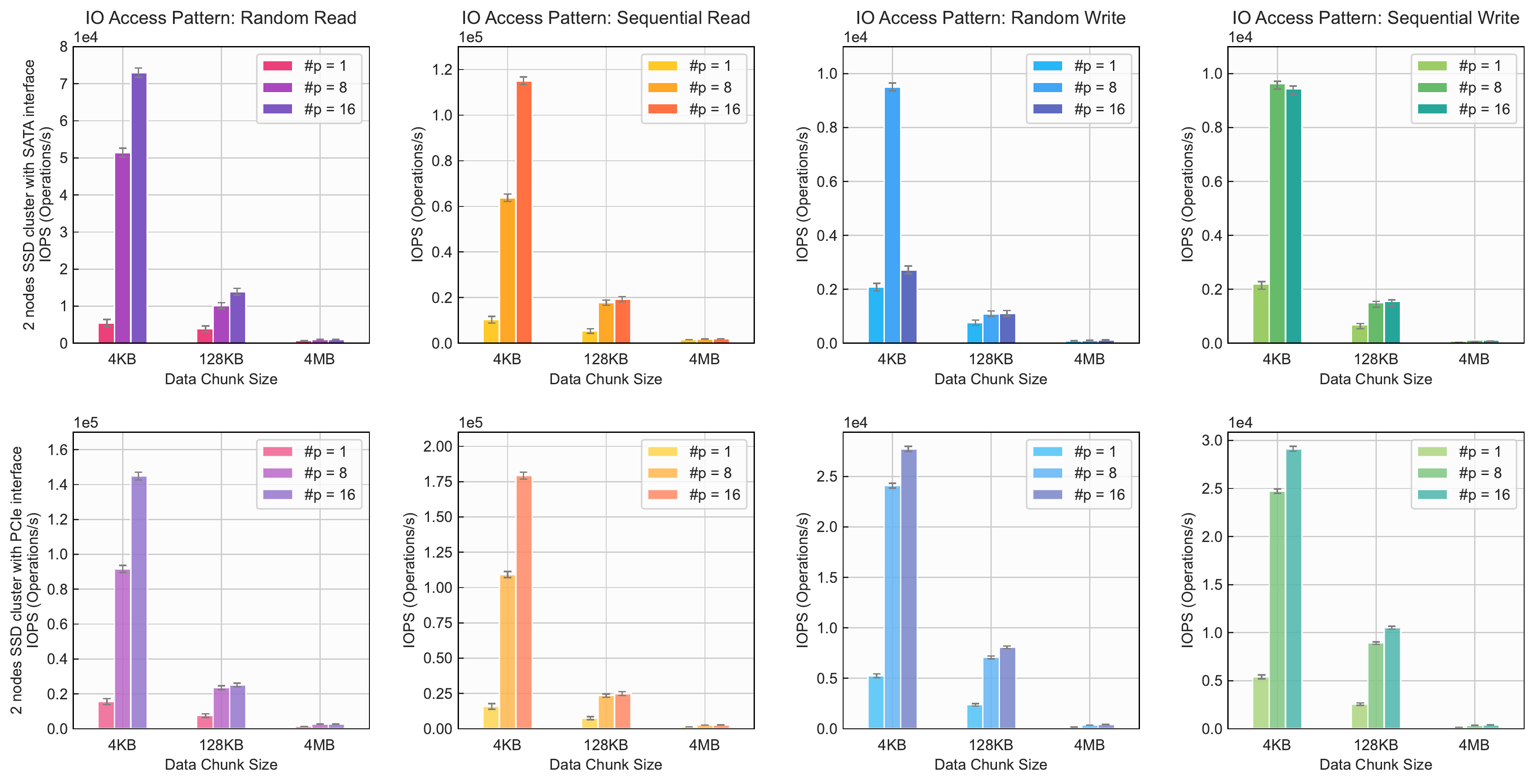}\\
    \caption{IOPS performance of the SSD cluster under different interface types (row in the figure), different I/O access patterns (column in the figure), different number of processes per client (represented in different colours), and different data chunk sizes (bar groups in the figure). $\#p$ represents number of client processes. \rw{The average results of 30 experiments are plotted, with corresponding error bars. }}
    \label{fig:figure4}
  \end{center}
\end{figure*}

\begin{figure*}[!ht]
  \begin{center}
    \includegraphics[width=\textwidth,keepaspectratio]{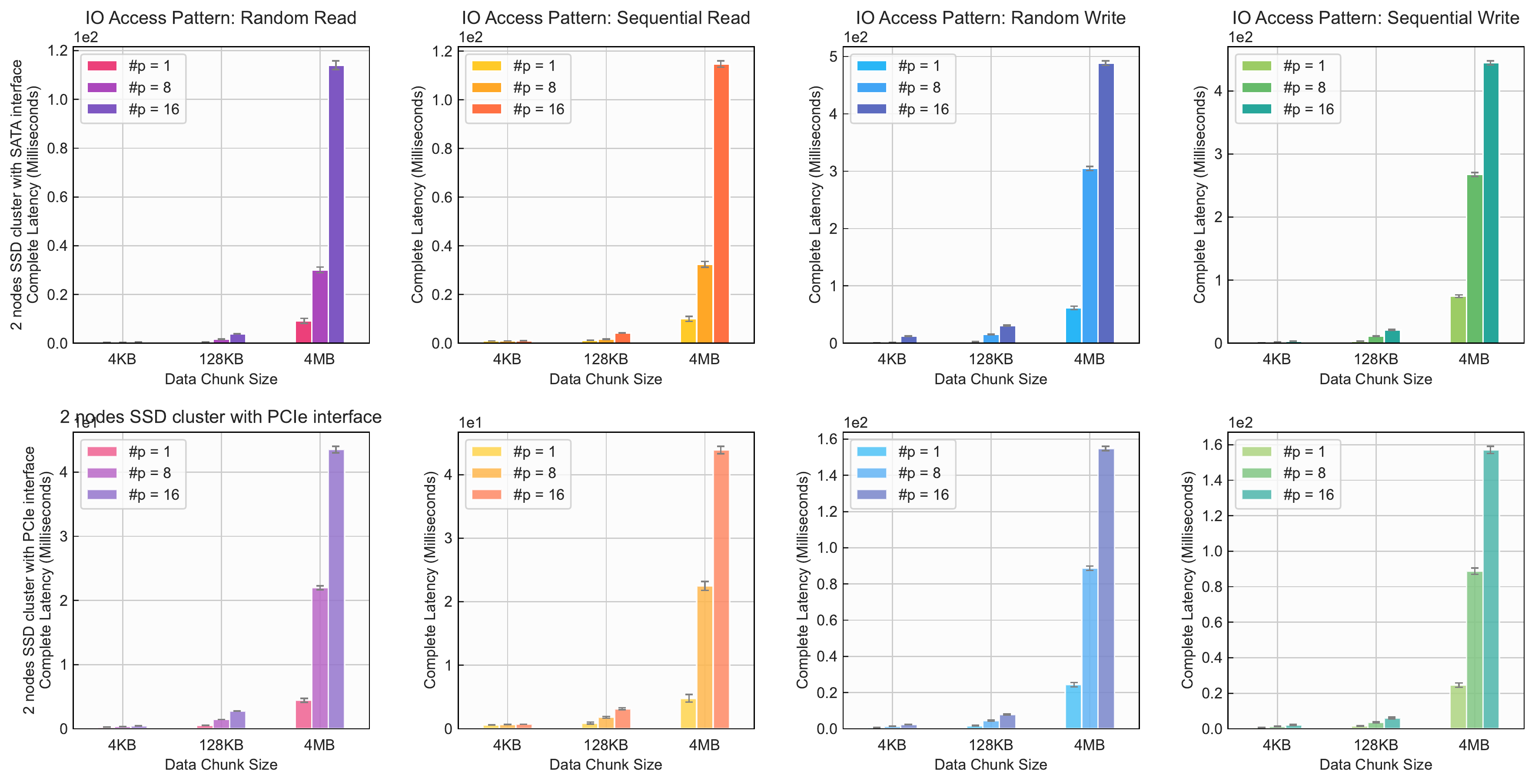}\\
    \caption{Complete Latency (clat) performance of the SSD cluster under different interface types (row in the figure), different I/O access patterns (column in the figure), different number of processes per client (represented in different colours), and different data chunk sizes (bar groups in the figure). $\#p$ represents number of client processes. \rw{The average results of 30 experiments are plotted, with corresponding error bars. }}
    \label{fig:figure5}
  \end{center}
\end{figure*}

We first evaluate the performance of the SSD cluster with $2$ storage nodes under different settings, including different SSD interface types, different I/O access patterns, different number of processes per client, and different data chunk sizes. Figure \ref{fig:figure3}, \ref{fig:figure4} and \ref{fig:figure5} visualise the evaluation results using bar charts for evaluation metrics of bandwidth, IOPS and complete latency, respectively. \rw{As aforementioned in Section \ref{sec:section4.1}, due to the superior performance of the PCIe SSD over SATA SSD, we do not use storage nodes with the same number of SSDs, instead, we reduce the number of SSDs in the PCIe SSD cluster node by half to make them relatively comparable. }

\subsubsection{Performance under different interface types}\label{sec:section5.1.1}

\rw{The comparison between different SSD interface types can be done by comparing two plots in each column in each figure (SATA on the top and PCIe on the bottom). }As we can observe, the relative height relationships of the bar charts and their trends are nearly the same for all three performance evaluation metrics under different SSD interface types. Hence, it indicates that in the SSD cluster with $2$ nodes, the trends of performance metrics are insensitive to SSD interface types, the SSD cluster will perform and scale in nearly the same way irrespective to the interface type of individual SSD that constitutes it. Thus, in later experiments, we are safe to analyse the performance of one SSD interface type, i.e., the more superior PCIe SSD. 

However, SSD interface types do significantly influence the performance of the SSD cluster, which is consistent with previous observations on the single SSD storage system \textsuperscript{\cite{krishnan2009integrating_pcie_outperform_sata}}. The bandwidth throughput of a 2-node PCIe SSD cluster is approximately $1.4$ to $3.2$ times higher than its SATA counterpart for random read, and its sequential read bandwidth is approximately $1.3$ to $1.9$ times higher than the 2-node SATA SSD cluster. The 2-node PCIe cluster further demonstrates its superiority to its SATA counterpart by achieving a $2.5$ to $10$ times higher bandwidth on random write and a $2.5$ to $7$ times higher bandwidth on sequential write, respectively. Moreover, the performance boosts with nearly the same magnitude of 2-node PCIe SSD cluster compared with its SATA counterpart on IOPS and a $8.8\%$ - $61.8\%$ significant latency drops are also observed, testifying the better bandwidth and IOPS performance and a lower latency of the 2-node PCIe SSD cluster, especially when performing write operations. Considering these observations, the 2-node SSD cluster with PCIe interface becomes a perfect solution for application scenarios in which write operations are frequently performed. 

\rw{\textbf{Deployment lessons: }Based on the above analyses, we gain the following lessons which can guide future decision-making: 
  \begin{itemize}
    \item The magnitude of performance boost under different settings and general trends are relatively the same for different SSD interface types. Hence, one can use the performance of SSD clusters of one interface type to approximately infer that of another SSD interface type, and therefore saves time. 
    \item Benefitting from the superiority of the PCIe SSD, the PCIe SSD cluster can achieve a significant performance boost. The bandwidth and the IOPS are significantly boosted and a lower latency is observed. 
    \item PCIe SSD clusters are more suitable to be used when write operations are frequently performed, since a more significant performance boost (PCIe over SATA) is observed on write operations, compared with its read counterparts. 
\end{itemize}}

\subsubsection{Performance under different I/O access patterns}\label{sec:section5.1.2}

In terms of the 2-node SSD cluster's performance under different I/O access patterns, investigations are done by comparing different columns in Figure \ref{fig:figure3} - \ref{fig:figure5}. For all three performance metrics, write operations' performance is inferior to read by a large margin for both interface types due to the overhead caused by SSD's write amplification and wear-leveling, etc., as is explained in Section \ref{sec:section2.1}. Quantitatively, it is interesting to observe that for the SATA SSD cluster, read operation's bandwidth and IOPS are all around $8$ - $12$ times higher than write operations, its latency achieves a $4$-time reduction compared with write operations. While for the PCIe SSD cluster, the gap between read and write performance is smaller than its SATA counterpart, with a performance improvement of around $2.2$ - $6.5$ times for bandwidth and IOPS, and the latency drops by around $2$ - $3.4$ times. The observations indicate that read and write I/O access patterns have a significant impact on the SSD cluster's performance, but the influence is less heavier on the SSD cluster with PCIe interface. By further analysing the results, we notice that the performance influence on the SSD cluster produced by random and sequential accesses is much lighter compared with the difference between read and write operations. The performance of bandwidth and IOPS under sequential I/O access is approximately $1.1$ - $1.5$ times better compared with the random I/O access, and the latency of the random I/O access is higher than sequential I/O access by only a small margin. This indicates that the SSD cluster's performance is relatively insensitive to the effect brought by random and sequential I/O access patterns thanks to its internal parallelism as introduced in Section \ref{sec:section2.1}, and is heavily affected by read and write operations due to the auxiliary operations such as write amplification and wear-leveling during writes as explained in Section \ref{sec:section2.1}. 

\textbf{Deployment lessons: }We learn the following lessons on SSD cluster I/O access patterns which can provide references for decision-making: 
\begin{itemize}
    \item The SSD cluster demonstrates a performance drop on write operations compared with read operations, due to the burdens associated with write operations such as wear-leveling, write amplifications, etc. 
    \item The PCIe SSD cluster experiences a less heavier performance drop on write operations compared with its SATA counterpart. Hence, it indicates that the PCIe SSD cluster not only enjoys superior overall performance, but also is more stable and relatively insensitive in terms of read/write operations. 
    \item Thanks to the internal parallelism of SSDs, for both interface types, sequential and random access patterns will not significantly influence cluster performances. 
\end{itemize}

\subsubsection{Performance scalability analysis under different number of client processes}\label{sec:section5.1.3}

As observed from Figure \ref{fig:figure3} to Figure \ref{fig:figure5}, besides an outlier which is the 2-node SATA SSD cluster under write I/O access, all other settings consistently show performance increase as the number of client processes increases, indicating that the SSD cluster scales relatively well when handling varied number of client processes. However, it is natural that no storage clusters can scale infinitely well, the performance increase in our SSD cluster is not linear in most cases as the number of client processes becomes larger. Take the PCIe SSD cluster as an example, under sequential read operations, increasing the number of client processes from $1$ to $8$ brings about $6.88$ times IOPS performance boost, while increasing the number of client processes from $8$ to $16$ only produces a $1.65$ times performance rise. Hence, it brings the network bandwidth bottleneck of the SSD cluster to our attention. When more client processes send the request simultaneously, the network bandwidth limitation between the client and the SSD cluster becomes the major bottleneck which limits the performance of the SSD cluster. This phenomenon will further be visualised in Figure \ref{fig:figure14} and analysed in Section \ref{sec:section5.5}. Therefore, when working with SSD storage clusters, it is worth investigating under current network capacity, which client process setting is the most suitable one to achieve a satisfying performance, as it is of less benefit to keep increasing the number of client processes due to the network bottleneck. 

\rw{\textbf{Deployment lessons: }In terms of the performance and the scalability of the SSD cluster when handling different number of client processes, we gain the following lessons: 
  \begin{itemize}
    \item The performance increases as the number of client processes increases. This is observed for both types of SSD clusters. But the performance increase is not proportional due to the presence of network bottleneck. 
    \item \rev{Due to the limitations such as the network bottleneck, the performance increase is not linear. Hence, when deploying the SSD storage clusters, one should bear in mind that the network bottleneck may exist and sending more client request processes may not receive a proportional performance boost. A moderate number of processes per client (8 in our case) is a good reference that can boost the performance of the SSD cluster while not encountering the network bottleneck. }
\end{itemize}}

\subsubsection{Performance and scalability analysis under different data chunk sizes}\label{sec:section5.1.4}

The SSD cluster's performances under read and write requests with different data chunk sizes have been evaluated by comparing bar groups in bar charts, as well as the trends present in Figure \ref{fig:figure3} to \ref{fig:figure5}. Observed from the Figure \ref{fig:figure3}, the SSD cluster's bandwidth performance shows a significant performance boost of more than three-fold when the data chunk size increases from $4$ KB to $128$ KB. \rw{However, due to the network bandwidth bottleneck as mentioned before, the increase of data chunk size from $128$ KB to $4$ MB doesn't proportionally bring the same magnitude of performance gain. The impact caused by the network bottleneck is clearly illustrated by the non-linear trend possessed by the bandwidth performance. }As for IOPS, it is natural to observe that the IOPS drops dramatically as the data chunk size increases, as shown in Figure \ref{fig:figure4}. Moreover, increasing the data chunk size dramatically rises the SSD cluster's latency according to Figure \ref{fig:figure5}. Hence, when leveraging the SSD cluster on latency-sensitive applications, suitable data chunk size should also be carefully chosen. In our case, set the data chunk size to be $128$KB will be a good choice, i.e., the bandwidth performance of the SSD cluster can be boosted without touching the network bottleneck, while not causing severe latency.

\rw{\textbf{Deployment lessons: }From the above analyses, we gain the following lessons in terms of data chunk size selection: 
  \begin{itemize}
    \item \rev{To fully exploit the efficiency of the SSD cluster, suitable data chunk size should be chosen such that the network bottleneck will not be reached. In our evaluation, $128$KB is a good choice to fully boost the bandwidth performance of the SSD cluster while not causing severe latency. Utilising a data chunk size that is too large can not only suffer from network bandwidth bottleneck, but also cause a low IOPS and a severe latency. }
    \item \rev{Utilising a data chunk size that is too large ($4$MB in our case) will drastically rise the latency, which is hurtful for latency-sensitive applications. }
\end{itemize}}

\begin{figure*}[!t]
  \begin{center}
    \includegraphics[width=0.75\textwidth,keepaspectratio]{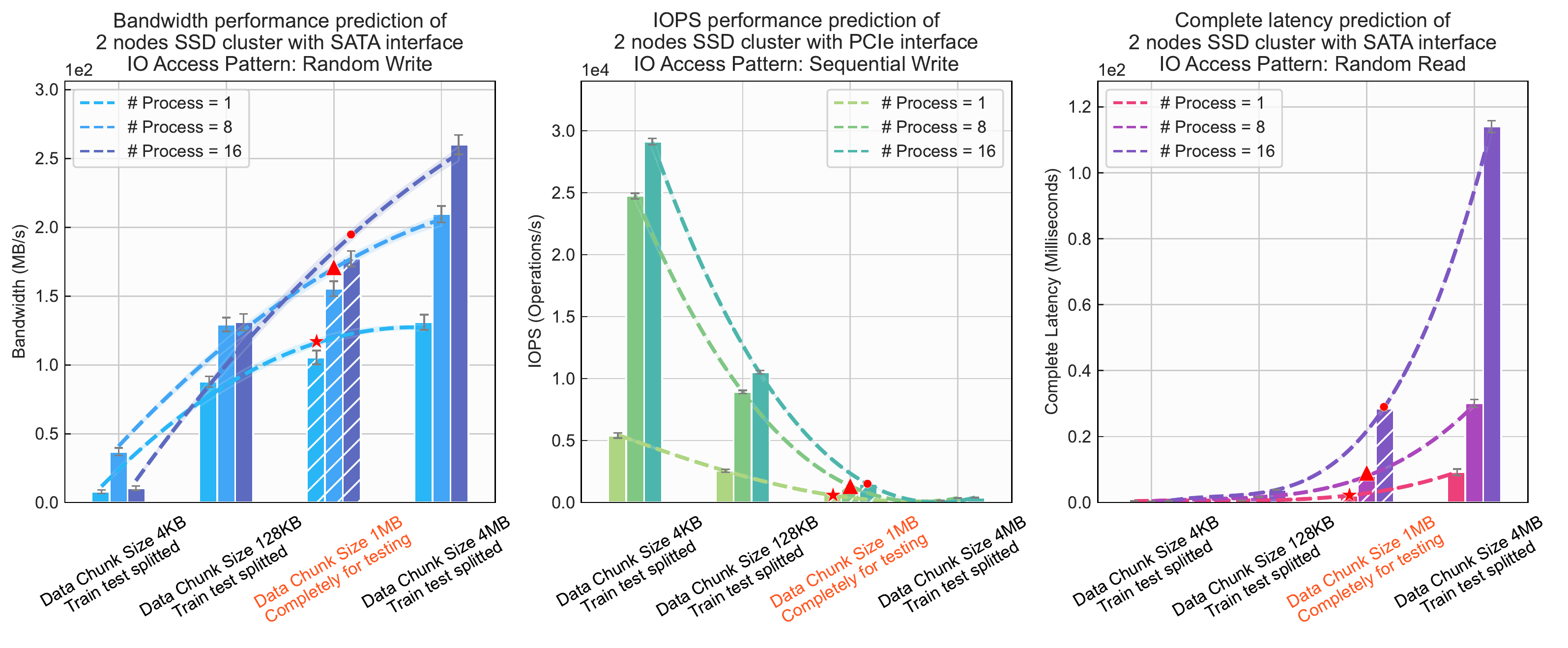}\\
    \caption{\revision{Qualitative visualisation of the generalisability of the fitted regression model on the 2-node SSD cluster under three randomly selected settings. }}
    \label{fig:figure6}
  \end{center}
\end{figure*}

\begin{table}
  \centering
  \revision{
  \caption{Quantitative evaluation results of the fitted regression models}
  \setlength{\tabcolsep}{2.0mm}{
    \begin{tabular}{c|cc}
      \toprule
      Number of processes per client & Root Mean Squared Error (RMSE) & R-square\Bstrut \\
      \hline    
      \multicolumn{3}{c}{Bandwidth RW}\Tstrut \\ 
      \hline
      Process = 1 & 18.17 & 0.79\Tstrut \\[0.3em]
      Process = 8 & 23.14 & 0.79 \\[0.3em]
      Process = 16 & 27.23 & 0.87\Bstrut \\
      \hline    
      \multicolumn{3}{c}{IOPS SW}\Tstrut \\ 
      \hline
      Process = 1 & 168.84 & 0.99\Tstrut \\[0.3em]
      Process = 8 & 584.31 & 0.99 \\[0.3em]
      Process = 16 & 735.55 & 0.99\Bstrut \\
      \hline    
      \multicolumn{3}{c}{Bandwidth RW}\Tstrut \\ 
      \hline
      Process = 1 & 0.98 & 0.88\Tstrut \\[0.3em]
      Process = 8 & 2.65 & 0.93 \\[0.3em]
      Process = 16 & 10.17 & 0.93\Bstrut \\
      \bottomrule
    \end{tabular}}
  \label{tab:table_regression}}
\end{table}

\begin{figure*}[!ht]
  \includegraphics[width=\textwidth,keepaspectratio]{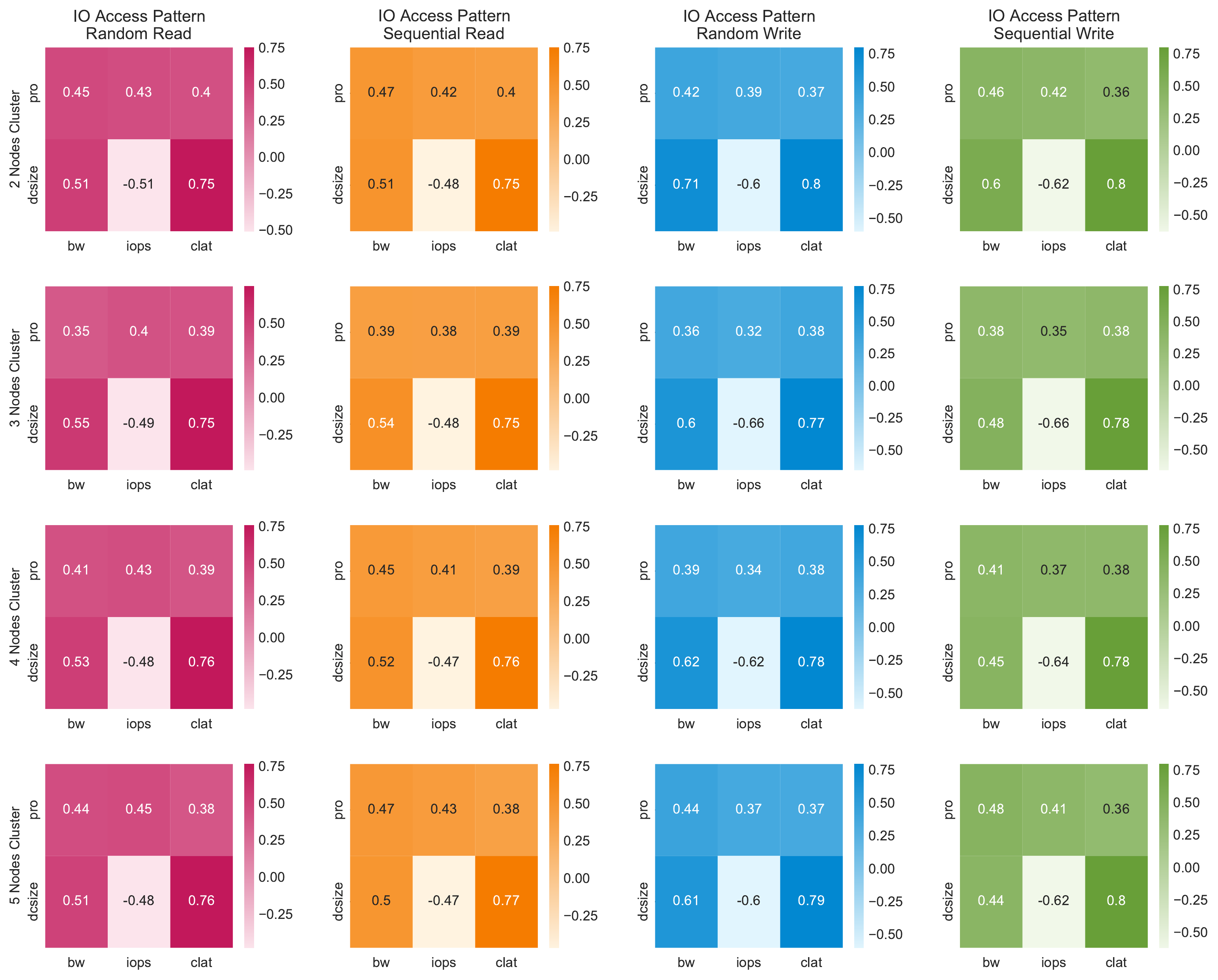}\\
  \caption{Heatmap between performance influential factors and performance metrics among different number of PCIe SSD cluster nodes under different I/O access patterns. }
  \label{fig:figure7}
\end{figure*}

\begin{figure*}[!ht]
  \begin{center}
    \includegraphics[width=0.8\textwidth,keepaspectratio]{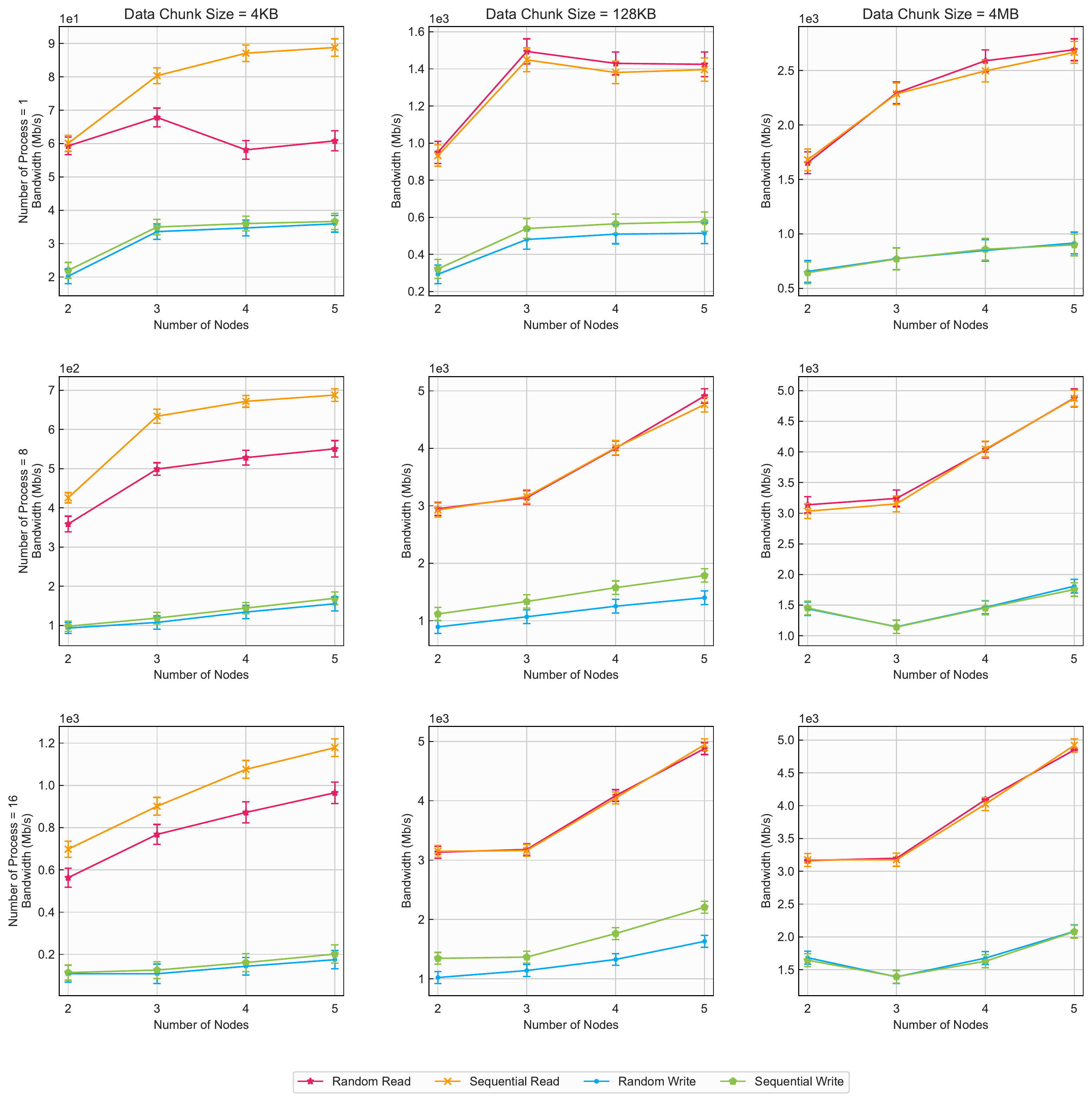}\\
    \caption{Bandwidth performance of the SSD cluster with different number of storage nodes under different number of processes per client (rows in the figure), different data chunk sizes (columns in the figure), and different I/O access patterns (represented by lines in different colours). \rw{The average results of 30 experiments are plotted, with corresponding error bars. }}
    \label{fig:figure8}
  \end{center}
\end{figure*}


\begin{figure*}[!ht]
  \begin{center}
    \includegraphics[width=0.8\textwidth,keepaspectratio]{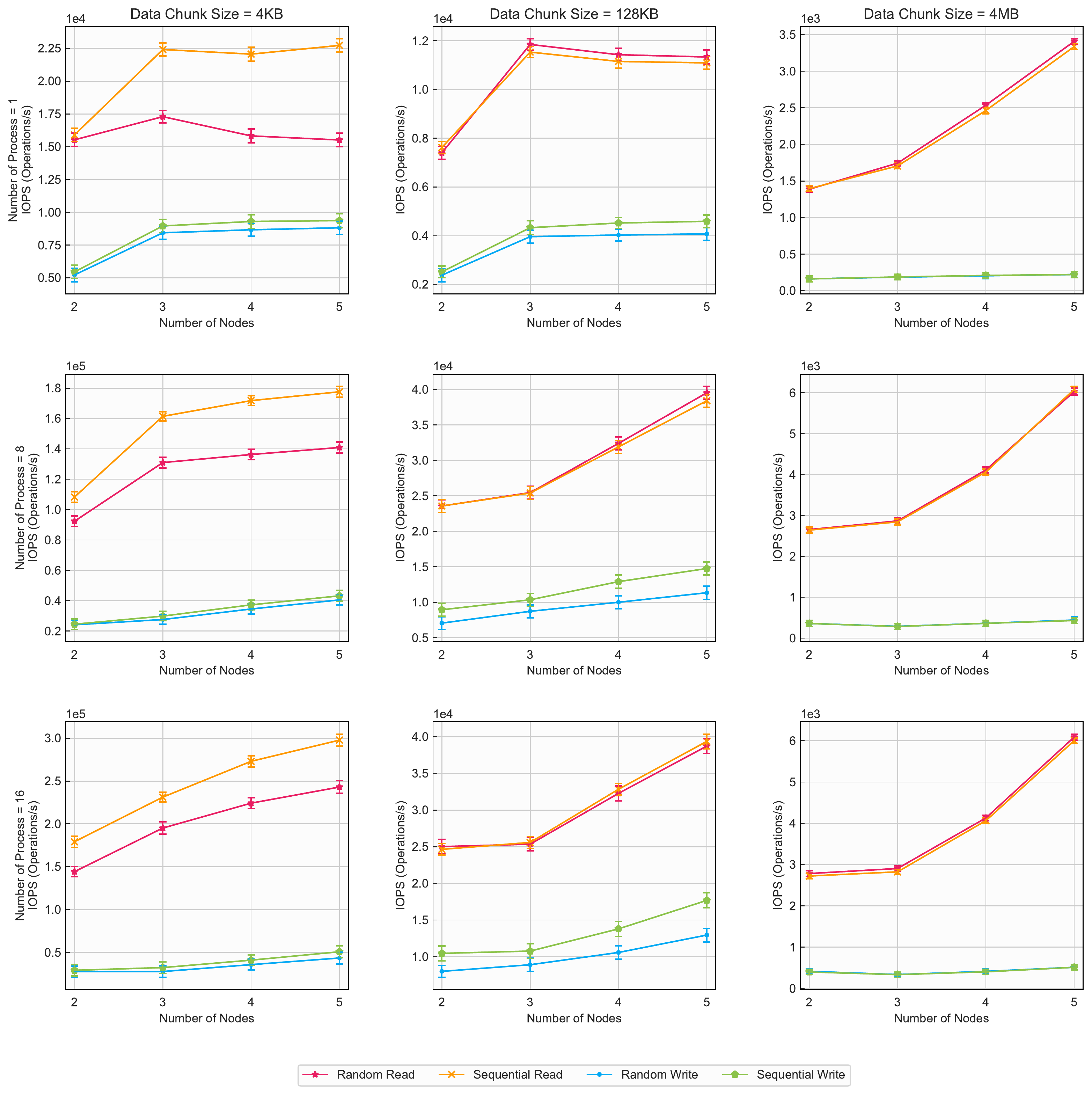}\\
    \caption{IOPS performance of the SSD cluster with different number of storage nodes under different number of processes per client (rows in the figure), different data chunk sizes (columns in the figure), and different I/O access patterns (represented by lines in different colours). \rw{The average results of 30 experiments are plotted, with corresponding error bars. }}
    \label{fig:figure9}
  \end{center}
\end{figure*}


\begin{figure*}[!ht]
  \begin{center}
    \includegraphics[width=0.8\textwidth,keepaspectratio]{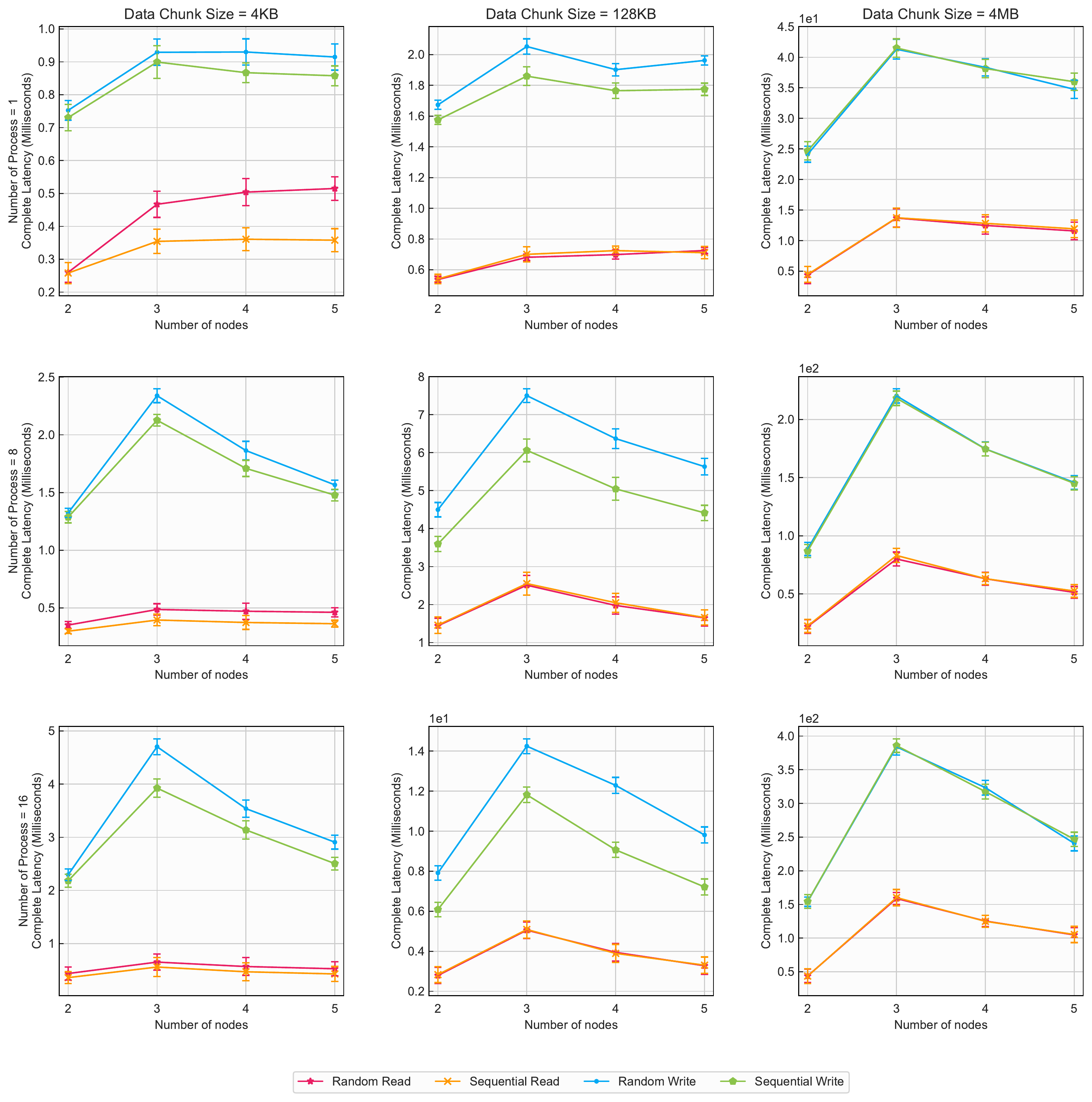}\\
    \caption{Complete latency performance of the SSD cluster with different number of storage nodes under different number of processes per client (rows in the figure), different data chunk sizes (columns in the figure), and different I/O access patterns (represented by lines in different colours). \rw{The average results of 30 experiments are plotted, with corresponding error bars. }}
    \label{fig:figure10}
  \end{center}
\end{figure*}

\subsection{\revision{2-node SSD cluster performance regression model analysis}\label{sec:section_regression}}

\revision{To make our performance evaluation more concrete and practical, and to further benefit the understanding and the applications of the SSD cluster, regression models have been fitted and evaluated, so that the results are predictive as much as possible on other configurations besides the ones tested. The visualisation of the fitted regression lines and their prediction results are illustrated in Figure \ref{fig:figure6} to qualitatively inspect its quality. Besides, quantitative evaluation results of regression lines in terms of their accuracy and goodness-of-fit are presented in Table \ref{tab:table_regression}. }

\revision{As shown in Figure \ref{fig:figure6}, the performance results data under $1$ MB data chunk size of three randomly selected settings have been presented and their bars in the bar chart have been marked with cross lines. The regression lines are fitted with the complete absence of data with 1MB data chunk size and the corresponding performance predictions yielded by the regression lines have been marked with red symbols. Qualitatively, the fitted regression lines' predictions are close to the real performance data, which indicate satisfying prediction capability. The regression line's capability is then analysed quantitative using rooted mean squared error (RMSE) and R-square as evaluation metrics. A relatively low RMSE indicates a satisfying fit, and R-square values close to 1 further verifies it, as the closer to 1 the R-square value is, the better the fit is. }

\revision{\textbf{Deployment lessons: }Therefore, by applying the regression model in the SSD cluster's performance analysis, it can guide the decision-making in the following ways: 
  \begin{itemize}
    \item \revision{We can use the fitted regression model to predict whether a certain setting can produce satisfying performance so that we can optimise the SSD cluster's configuration settings without explicitly testing it. }
    \item \revision{By comparing the performance predictions produced by the fitted regression model and the monitored performance, we can then perform fault detection to detect whether the SSD cluster is functioning properly. Significant deviations between the predicted performance and the monitored performance may imply that there may exist some issues in the SSD cluster and special attention is needed so that the SSD cluster can keep functioning properly and efficiently. }
\end{itemize}}

\subsection{Performance influential factors effectiveness under different number of storage nodes}\label{sec:section5.3}

As is explained in Section \ref{sec:section5.1.1}, the rest of the experiments are conducted on the PCIe SSD cluster. To fully explore the performance of the SSD cluster, the SSD cluster with not only 2 storage nodes, but also with more than 2 nodes, have been experimented and evaluated. It is natural to investigate whether performance influential factors including different I/O access patterns, different number of processes per client and different data chunk sizes, will affect the SSD cluster's performance in the same way when the cluster possesses different number of nodes. Hence, it leads to the motivation of having Figure \ref{fig:figure7}. In Figure \ref{fig:figure7}, heatmaps between performance influential factors and performance metrics under different settings have been presented. In each heatmap, the number indicates the correlation coefficient between the influential factor and the performance metric, the colour visualises the degree of influence. The deeper the colour is, the more positively correlated relationship it indicates, and the lighter the colour is, the more negatively correlated relationship it indicates. 

\rw{By comparing each row in Figure \ref{fig:figure7}, we notice that all performance influential factors are affecting the SSD cluster's performance metrics in the same manner and to nearly the same degree when the number of nodes in the SSD cluster varies, as specified by the same visual looking of heatmap's colour. }This reveals relatively good scalability of the SSD cluster. The number of client processes ($pro$) positively correlates with all performance metrics under all I/O access patterns with nearly the same magnitude when the number of storage nodes in the SSD cluster varies. The data chunk size ($dcsize$) also positively correlates with bandwidth performance of SSD cluster, second to the complete latency, which heavily correlates with data chunk size with a correlation coefficient around $0.75$ - $0.8$ in all settings. The IOPS on the other hand negatively correlates with data chunk sizes in all cases by having a correlation coefficient around $-0.66$ - $-0.48$. Hence, based on this observation, it reveals good scalability of the SSD cluster and hence, it alleviates the necessity of analysing the performance of the SSD cluster with different number of storage nodes in great details as is previously done. 

\rw{\textbf{Deployment lessons: }From the performance influential factor and performance metric correlation analysis, we learned the following lessons: 
  \begin{itemize}
    \item The SSD cluster scales well, different performance influential factors influence the performance in the same way with nearly the same degree when the number of storage nodes varies. 
    \item This excellent scalability makes the results representative and eases further performance evaluation process. It alleviates the necessity of analysing the performance of the SSD cluster with different number of storage nodes in great details. 
\end{itemize}}

\begin{figure*}[!ht]
  \begin{center}
    \includegraphics[width=0.8\textwidth,keepaspectratio]{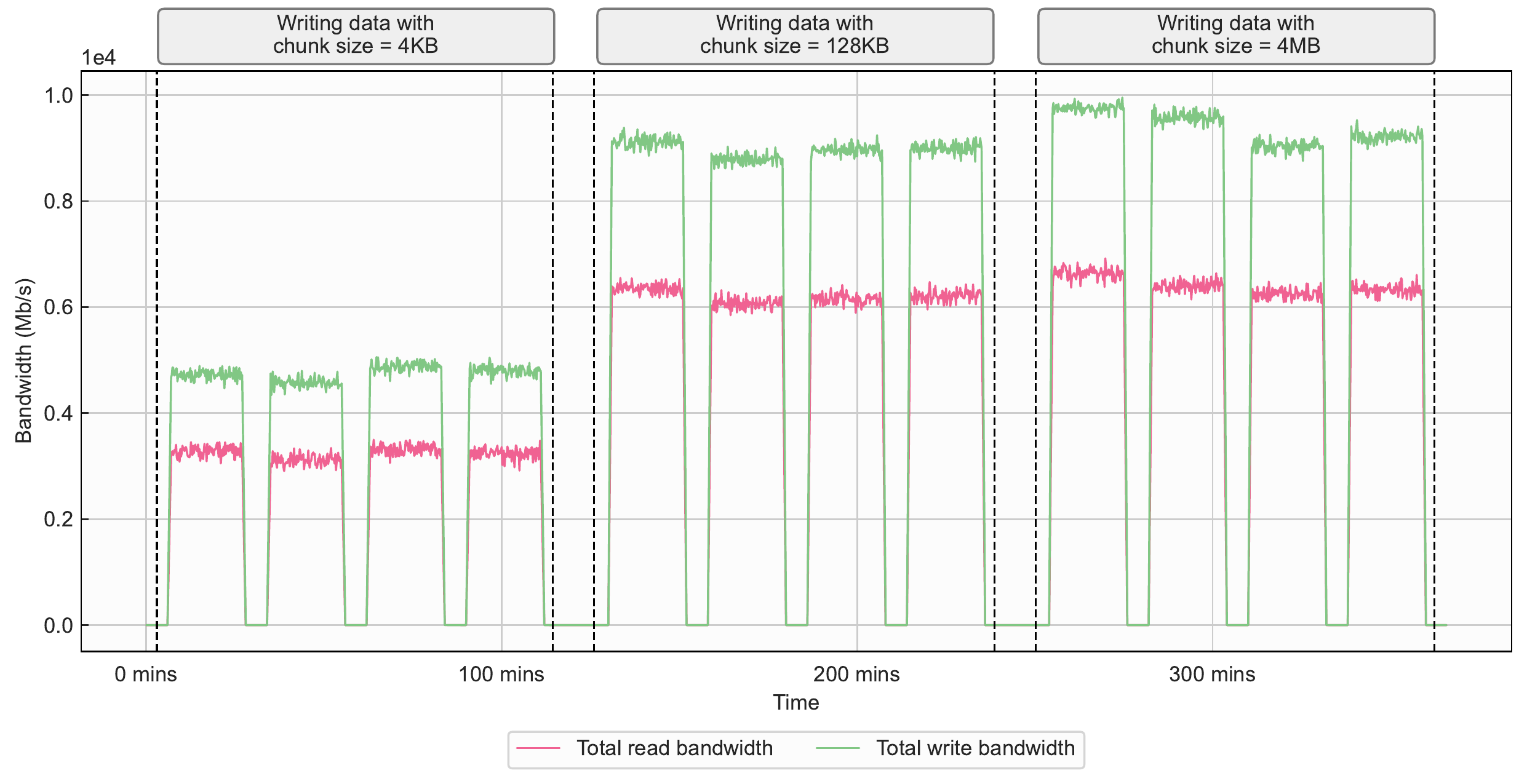}\\
    \caption{\rev{Network traffic measured in bandwidth (Mb/s), monitored when writing data with different data chunk sizes. The randomly selected setting is a $4$-node SSD cluster with the PCIe interface, the I/O access pattern is random write, with $16$ processes per client. Each write operation is repeated $4$ times and all results are plotted. }}
    \label{fig:figure14}
  \end{center}
\end{figure*}

\subsection{Performance of SSD cluster with multiple nodes}\label{sec:section5.4}

We now investigate the performance of the SSD cluster with different number of storage nodes under settings (Table \ref{tab:table3}) including different number of processes per client, different data chunk sizes and different I/O access patterns in Figure \ref{fig:figure8} to \ref{fig:figure10}. By observing Figure \ref{fig:figure8}, we notice that as the number of storage nodes increases, the SSD cluster's bandwidth performance shows a generally rising trend in all settings. It is also interesting to note that the increase of data chunk size and number of processes per client doesn't hugely impact the performance increasing rate when changing the number of storage nodes in SSD cluster from $2$ to $5$. Except for the top-left case that does not fully exploit internal parallelism of the cluster due to minimum data chunk sizes and number of processes per client, all other settings show an approximately $50\%$ - $60\%$ bandwidth performance increase when we rise the number of storage nodes from $2$ to $5$. This suggests that the performance boost ratio of the SSD cluster is relatively insensitive to the changes of data chunk size and number of processes per client for SSD clusters with different number of nodes, and increasing the number of storage nodes in the SSD cluster can stably improve its bandwidth performance, which also justifies the excellent scalability of the SSD cluster. 

As for SSD cluster's IOPS performance as shown in Figure \ref{fig:figure9}, its trend presents nearly the same manner as the bandwidth. Finally, in terms of the complete latency of SSD cluster with different number of storage nodes as illustrated in Figure \ref{fig:figure10}, the latency keeps fluctuating without showing clear trends or peaks, and more specifically, the fluctuation is within the reasonable range of around $20\%$. Hence, this observation further demonstrates the excellent scalability and high efficiency of the SSD cluster. 

\rev{Upon closer inspection of Figure \ref{fig:figure8} to \ref{fig:figure10}, we can observe that when the data chunk size is relatively small, i.e., $4$ KB, there is a performance gap between random and sequential accesses, especially when the SSD cluster possesses more storage nodes. When the data chunk size becomes $128$KB, the random and sequential read access present no significant difference under medium data chunk size setting, while write operations are quite a different story. Finally, when setting the data chunk size to be $4$MB which is relatively large, random and sequential accesses for both read and write operations don't present significant performance gaps at all, which reveals that the SSD cluster is insensitive to random or sequential access patterns when the data chunk size is relatively large. }This is due to the internal parallelism mechanism of the SSD as mentioned in Section \ref{sec:section2.1}. When the data chunk size aligns with the size of the clustered block, data will be stripped over multiple channels which fully exploits the internal parallelism, and hence, random and sequential access patterns present no significant performance gaps. By observing this in SSD clusters, it demonstrates that when setting data chunk size to be relatively large, SSD cluster is suitable to process data regardless of whether its I/O access pattern is sequential or not. 

\rw{\textbf{Deployment lessons: }Based on the above analyses, we gain the following knowledge: 
  \begin{itemize}
    \item The SSD storage cluster scales well when the number of storage nodes increases. The performance trend is relatively insensitive to the changes of data chunk size and number of processes per client. 
    \item Under small data chunk size, the SSD storage cluster will present a better performance on sequential read/write operations than random accesses. Under medium data chunk size, the SSD storage cluster tends to be insensitive on read accesses, but for the more complex write accesses, the gap between random and sequential access performance still exists. By fully exploiting the SSD's internal parallelism mechanism, the SSD cluster with relatively large data chunk sizes is insensitive to random or sequential I/O access patterns. 
\end{itemize}}

\subsection{Illustration and analysis of the network bandwidth bottleneck}\label{sec:section5.5}

To visualise the network bandwidth bottleneck encountered during experiments, we randomly select a representative case to highlight the network bottleneck. Figure \ref{fig:figure14} presents the network traffic that we monitored during random write I/O access on a 4-node PCIe SSD cluster, with $16$ processes per client and varied data chunk sizes. \rw{Each write operation is repeated $4$ times and all results are plotted. The network configuration has been presented in Section \ref{sec:section4.1}. }

The first, second and third peak groups in Figure \ref{fig:figure14} shows the network traffic when writing data with chunk size set to $4$ KB, $128$ KB and $4$ MB, respectively. Without the presence of the network bandwidth bottleneck, the write operation's bandwidth should grow relatively proportionally as the data chunk size grows. As we can observe, when writing data chunk with size $4$ MB, the network bandwidth remains nearly the same as when writing data with data chunk size $128$ KB, and this is where the network bandwidth bottleneck occurs. Due to the existence of network bandwidth bottleneck, the capability of the SSD cluster cannot be fully excavated, causing the mediocre performance even though we utilise more processes and large data chunk sizes, and hence, it makes these attempts fruitless. 

\rw{\textbf{Deployment lessons: }Based on the above experiments, we learn the following lessons: 
  \begin{itemize}
    \item The presence of the network bandwidth bottleneck may cause the capability of the SSD cluster failed to be fully excavated. 
    \item \rev{During deployment, when adjusting a performance influencing factor, e.g., the number of client processes, if the performance does not increase accordingly, then the network bottleneck may be a cause. }
\end{itemize}}


\section{Conclusion}\label{sec:section6}

In this study, we evaluate the performance of the SSD cluster rather than focusing on a single SSD. We evaluate various performance metrics under different settings, including different number of storage nodes in the SSD cluster, different interface types, different I/O access patterns, different data chunk sizes, and different number of processes per client, which are comprehensive. The performances and their trends are comprehensively compared and evaluated using analytical tools to make the results concrete and persuasive. \revision{We show that the SSD cluster exhibits exceptional performance, build a regression model to facilitate performance prediction and anomaly detection, and testify the excellent scalability of the SSD cluster with respect to influential factors. }The network bottleneck observed during experiments is also analysed and discussed. \rw{We summarise the lessons and knowledge we gained from experiments to guide decision-making in future cluster deployment. We believe that our exploration can inspire more efforts in the utilisation of SSD clusters. }

\section*{Acknowledgement}
\label{sec:section_acknowledgement}

This work is supported by the Third Xinjiang Scientific Expedition Program (Grant No.2021xjkk1300). 

\section*{Data Availability Statement}
\label{sec:section_data_availability_statement}

The data that support the findings of this study are available from the corresponding author upon reasonable request. 

\begin{spacing}{0.85}
\bibliography{SSD}
\end{spacing}

\end{document}